**Colossal barocaloric effects near room temperature in plastic crystals of neopentylglycol**

P. Lloveras[1], A. Aznar[1], M. Barrio[1], Ph. Negrier[2], C. Popescu[3], A. Planes[4], L. Mañosa[4], E. Stern-Taulats[5], A. Avramenko[5], N. D. Mathur[5], X. Moya[5,*] and J.-Ll. Tamarit[1,*]

[1]Grup de Caracterització de Materials, Departament de Física, EEBE and Barcelona Research Center in Multiscale Science and Engineering, Universitat Politècnica de Catalunya, Eduard Maristany, 10-14, 08019 Barcelona, Catalonia

[2]Université de Bordeaux, LOMA, UMR 5798, F-33400 Talence, France

[3]CELLS-ALBA Synchrotron, E-08290 Cerdanyola del Vallès, Catalonia, Spain

[4]Departament de Física de la Matèria Condensada, Facultat de Física, Universitat de Barcelona, Martí i Franquès 1, 08028 Barcelona, Catalonia

[5]Department of Materials Science, University of Cambridge, Cambridge, CB3 0FS, UK

e-mail: xm212@cam.ac.uk, josep.lluis.tamarit@upc.edu

**There is currently great interest in replacing the harmful volatile hydrofluorocarbon fluids used in refrigeration and air-conditioning with solid materials that display magnetocaloric, electrocaloric or mechanocaloric effects. However, the field-driven thermal changes in all of these caloric materials fall short with respect to their fluid counterparts. Here we show that plastic crystals of neopentylglycol $(CH_3)_2C(CH_2OH)_2$ display unprecedentedly large pressure-driven thermal changes near room temperature due to molecular reconfiguration, and that these changes are comparable with those exploited commercially in hydrofluorocarbons. Our discovery of colossal barocaloric effects in a plastic crystal should bring barocaloric materials to the forefront of research and development in order to achieve safe environmentally friendly cooling without compromising performance.**

Plastic crystals (PCs), also known as orientationally disordered crystals, are materials that lie at the boundary between solids and liquids[1]. They are normally made of nearly spherical small organic molecules whose centres of mass form a regular crystalline lattice[1], unlike liquid crystals that normally comprise highly anisotropic organic molecules with no long-range positional order[2]. The globular shape of these molecules provides little steric hindrance for reorientational processes, such that plastic crystals tend to be highly orientationally disordered away from low temperature[3]. This dynamical disordering often implies high plasticity under uniaxial stress, and hence the materials are known as plastic crystals[4]. On cooling, plastic crystals typically transform into ordered crystals (OCs) of lower volume via first-order phase transitions, whose latent heats arise primarily due to thermally driven large changes of orientational order, and this has led to proposals for passive thermal storage[5,6]. Here we exploit commercially available samples of the prototypical plastic crystal neopentylglycol (NPG), i.e. 2,2-dymethyl-1,3-propanediol. This material is an alcoholic derivative of neopentane $C(CH_3)_4$ made from cheap abundant elements, and enjoys widespread use in industry as an additive in the synthesis of paints, lubricants and cosmetics.

We achieve colossal pressure-driven thermal changes (barocaloric effects) near room temperature that are an order of magnitude better than those observed in state-of-the-art barocaloric (BC) materials[7-17] and comparable to those observed in the standard commercial hydrofluorocarbon refrigerant R134a[18] (Table 1). Our BC effects are colossal because the first-order PC-OC transition displays an enormous latent heat that is accompanied by an enormous change in volume, such that moderate applied pressure is sufficient to yield colossal thermal changes via the reconfiguration of globular neopentylglycol molecules (whose steric hindrance is low[3]). Moreover, reversibility is achieved at temperatures above the hysteretic transition regime. Our higher operating pressures do not represent a barrier for applications because they can be generated by a small load in a large volume of material via a pressure-transmitting medium, e.g. using a vessel with a neck containing a driving piston, whose small area is compensated by its distance of travel. Therefore, our demonstration of colossal BC effects in commercially available plastic crystals should immediately open new avenues for the development of safe and environmentally friendly solid-state refrigerants.

At room temperature and atmospheric pressure, NPG adopts an ordered monoclinic structure ($P2_1/c$) with four molecules per unit cell[19] [Figure 1(a)]. On heating, the material undergoes a reversible structural phase transition to a cubic structure ($Fm\bar{3}m$) with four molecules per unit cell that adopt an orientationally disordered configuration at any typical instant[20] [Figure 1(a)]. The first-order structural phase transition yields sharp peaks in $dQ/|dT|$ ($Q$ is heat, $T$ is temperature) recorded on heating and cooling [Figure 1(a)], with a well-defined transition start temperature $T_0 \sim 314$ K on heating (Figure S1). By contrast, as a consequence of the nominally isothermal character of the PC-OC transition[21], the temperature ramp rate influences the transition finish temperature on heating, and the transition start and finish temperatures on cooling (e.g. by up to ~5 K for 1-10 K min$^{-1}$, Figure S1). Integration of the calorimetric peaks yields a large latent heat of $|Q_0| = 121 \pm 2$ kJ kg$^{-1}$ on heating, and $|Q_0| = 110 \pm 2$ kJ kg$^{-1}$ on cooling [Figure 1(a)]. These values of $|Q_0|$ are independent of the temperature ramp rate (Figure S1), and in good agreement with previous experimental values[1,22,23] of $|Q_0| \sim 123$-$131$ kJ kg$^{-1}$.

Integration of $(dQ/|dT|)/T$ and $C_p/T$ [Figure 1(b)], permits the evaluation of entropy $S'(T) = S(T) - S(250\text{ K})$ over a wide temperature range [Figure 1(c)], as explained in the Experimental Section ($C_p$ is specific heat at atmospheric pressure). The large entropy change at the transition ($|\Delta S_0| \sim 383$ J K$^{-1}$ kg$^{-1}$ on heating and $|\Delta S_0| \sim 361$ J K$^{-1}$ kg$^{-1}$ on cooling) is in good agreement with previous experimental values[1,21-23] of $|\Delta S_0| \sim 390$-$413$ J K$^{-1}$ kg$^{-1}$. This large value of |$\Delta S_0$| arises due to a non-isochoric order-disorder transition in molecular configurations, such that it exceeds values of |$\Delta S_0$| << 100 J K$^{-1}$ kg$^{-1}$ for first-order structural phase transitions associated with changes of ionic position[24-27] and electronic densities of states[24,27,28]. Consequently, the configurational degrees of freedom that are accessed via the non-isochoric order-disorder transition in our solid material yield entropy changes that compare favourably with those associated with the translational degrees of freedom accessed via solid-liquid-gas transitions in various materials[29], including the hydrocarbon fluids used for commercial refrigeration[18].

On heating through the transition, x-ray diffraction data confirm the expected changes in crystal structure[19,20] (Figure S6 and S7). The resulting specific volume $V$ undergoes a large ~4.9% increase of $\Delta V_0$ = 0.046 ± 0.001 cm$^3$ g$^{-1}$ across the transition, for which $(\partial V/\partial T)_{p=0} > 0$ [Figure 1(d)], presaging large conventional BC effects that may be evaluated[30] by using the Maxwell relation $(\partial V/\partial T)_p = -(\partial S/\partial p)_T$ to calculate the isothermal entropy change $\Delta S(p_1 \to p_2) = -\int_{p_1}^{p_2} (\partial V/\partial T)_p \, dp$ due to a change in pressure from $p_1$ to $p_2$. Near the transition, the volumetric thermal expansion coefficients for the OC and the PC phases are both ~10$^{-4}$ K$^{-1}$, implying the existence of additional[15] BC effects $\Delta S_+$ that are large and conventional at temperatures lying on either side of the transition. These additional BC effects are evaluated here using the aforementioned Maxwell relation, for changes in pressure |$p - p_{atm}$| ~ |$p$| where atmospheric pressure $p_{atm}$ ~ 0 GPa, to obtain $\Delta S_+(p) = -[(\partial V/\partial T)_{p=0}]p$, where $(\partial V/\partial T)_p$ is assumed to be independent of pressure[13,15,17] (Figure S4 shows the error in $(\partial V/\partial T)_p$ to be ~20% for the PC phase, which implies an error in the total entropy change $\Delta S$ of ~3%).

Two contributions to $|\Delta S_0|$ may be identified as follows. One is the configurational entropy[31,32] $M^{-1} R \ln \Omega$, where $M = 104.148$ g mol$^{-1}$ is molar mass, $R$ is the universal gas constant, and $\Omega$ is the ratio between the number of configurations in the PC and the OC $\alpha\kappa)\Delta V_0$, where the coefficient of isobaric thermal expansion $\alpha$ (Figure S4), and the isothermal $\kappa$ (Figure S5), have both been averaged across the PC-OC transition. Molecules of $(CH_3)_2C(CH_2OH)_2$ display achiral tetrahedral symmetry[33] (point group Td, subgroup C3v), yielding one configuration in the OC phase and 60 configurations in the PC phase (10 molecular orientations that each possesses six possible hydroxymethyl conformations). Therefore the configurational entropy is $M^{-1} R \ln 60 \sim 330$ J K$^{-1}$ kg$^{-1}$, and the volumetric entropy is $\sim 60$ J K$^{-1}$ kg$^{-1}$ [data from Figure 1(d) and Figure S3(a)]. The resulting prediction of $|\Delta S_0| \sim 390$ J K$^{-1}$ kg$^{-1}$ agrees well with the experimental values reported above, and the previously measured experimental values[1,21-23] reported above.

Measurements of $dQ/|dT|$ under applied pressure [Figure 2(a,b)] reveal that the observed transition temperatures vary strongly with pressure [Figure 2(c)], with $dT/dp = 113 \pm 5$ K GPa$^{-1}$ for the start temperature on heating, and $dT/dp = 93 \pm 18$ K GPa$^{-1}$ for the start temperature on cooling, for pressures $p < 0.1$ GPa [black lines, Figure 2(c)]. These values of $dT/dp$ are amongst the largest observed for BC materials (Table S1), and indicate that the first-order PC-OC transition of width $\sim 10$ K [Figure 2(a,b)] could be fully driven in either direction using $|\Delta p| \sim |p| \sim 0.1$ GPa. At higher pressures, values of $dT/dp$ fall slightly, but remain large [Figure 2(c)].

Integration of $(dQ/|dT|)/T$ at finite pressure reveals that the entropy change $|\Delta S_0|$ decreases slightly with increasing pressure [Figure 2(d)]. This decrease arises because the additional entropy change $\Delta S_+(p)$ increases in magnitude on increasing temperature in the PC phase [$(\partial V/\partial T)_{p=0}$ at 370 K is ~240% larger than $(\partial V/\partial T)_{p=0}$ at 320 K, Figure 1(d)], whereas it is nominally independent of temperature in the OC phase near the transition. The fall seen in both d$T$/d$p$ and |$\Delta S_0$| implies via the Clausius-Clapeyron equation d$T$/d$p$ = $\Delta V_0/\Delta S_0$ that there is a reduction in |$\Delta V_0$| at finite pressure [Figure 2(e)], as confirmed using pressure-dependent dilatometry [Figure S3(a)] and pressure-dependent x-ray diffraction [Figure S3(b)].

In order to plot $\Delta S(T,p)$, we obtained finite-pressure plots of $S'(T,p) = S(T,p) - S(250 \text{ K},0)$ [Figure 3(a,b)] by integrating the data in Figure 2(a,b) and Figure 1(b), and displacing each corresponding plot by $\Delta S_+(p)$ at 250 K, as explained in the Experimental Section. (Note that $\Delta S_+(p)$ was evaluated below $T_0(p=0)$ to avoid the forbidden possibility of $T_0(p)$ rising to the temperature at which $\Delta S_+(p)$ was evaluated at high pressure.) From Fig. 3(a,b), we see that the entropy change associated with the transition $\Delta S_0(p)$ combines with the smaller same-sign additional entropy change $\Delta S_+(p)$ away from the transition, yielding total entropy change $\Delta S(p)$.

By following isothermal trajectories in our plots of $S'(T,p)$ obtained on cooling [Figure 3(b)], we were able to evaluate $\Delta S(T,p)$ on applying pressure [Figure 3(c)], as cooling and high pressure both tend to favour the low-temperature low-volume OC phase. Similarly, by following isothermal trajectories in our plots of $S'(T,p)$ obtained on heating [Figure 3(a)], we were able to evaluate $\Delta S(T,p)$ on decreasing pressure [Figure 4(c)], as heating and low pressure both tend to favour the high-temperature high-volume PC phase.

Discrepancies in the magnitude of $\Delta S(T,p)$ on applying and removing pressure [Figure 3(c)] are absent in the range ~314-342 K, evidencing reversibility. Our largest reversible isothermal entropy change |$\Delta S$| ~ 510 J K$^{-1}$ kg$^{-1}$ arises at ~320 K for |$p$| ~ 0.52 GPa, and substantially

exceeds the BC effects of $|\Delta S| \leq 70$ J K$^{-1}$ kg$^{-1}$ that were achieved using similar values of $|p|$ in a range of materials near room temperature [Figure 4(a)], namely magnetic alloys[7-12,34], ferroelectric[13,35,36] and ferrielectric[15] materials, fluorides and oxifluorides[14,37-40], hybrid perovskites[16], and superionic conductors[17,41,42]. Moreover, our largest value of $|\Delta S|$ substantially exceeds the values recorded for magnetocaloric[30,43-46], electrocaloric[30,47,48], and elastocaloric[30,49] materials, and is comparable to the values observed in the standard commercial hydrofluorocarbon refrigerant fluid R134a[18], for which $|\Delta S| = 520$ J K$^{-1}$ kg$^{-1}$ at ~310 K for much smaller operating pressures of ~ 0.001 GPa [Figure 4(a)]. We can also confirm that NPG compares favourably with other BC solids[7,10,8,11,9,12,31] when normalizing the peak entropy change by volume[30] to yield $|\Delta S| \sim 0.54$ J K$^{-1}$ cm$^{-3}$ (the NPG density is 1064 kg m$^{-3}$ at ~320 K).

The large variation of transition temperature with pressure [Figure 2(c)] permits large entropy changes of $|\Delta S| \sim 445$ J K$^{-1}$ kg$^{-1}$ to be driven with relatively moderate pressure changes of $|p| \sim 0.25$ GPa [Figure 3(c)], yielding giant BC strengths[30] of $|\Delta S|/|p| \sim 1780$ J K$^{-1}$ kg$^{-1}$ GPa$^{-1}$. Larger pressures extend the reversible BC effects to higher temperatures [Figure 3(c)], causing the large refrigerant capacity RC to increase [Figure 4(b)] despite the slight reduction in $|\Delta S_0(p)|$ [Figure 2(d)]. The BC effects in NPG are so large [Figure 4(a)] that unpractical changes of pressure would be required to achieve comparable RC values in other BC materials.

By following adiabatic trajectories in $S'(T,p)$ [Figure 3(a,b)], we established both the adiabatic temperature change $\Delta T(T_s,p)$ on applying pressure $p$ at starting temperature $T_s$ [Figure 3(d)], and the adiabatic temperature change $\Delta T(T_f,p)$ on removing pressure $p$ to reach finishing temperature $T_f$ [Figure 3(e)]. On applying our largest pressure ($p \sim 0.57$ GPa), an adiabatic temperature increase of $\Delta T \sim 30$ K with respect to $T_s \sim 318$ K is necessarily reversible above the thermally hysteretic regime, such that an equivalent temperature change of opposite sign is achieved on pressure removal. These BC effects substantially exceed both the BC effects of $|\Delta T| \leq 10$ K that were achieved in inorganic materials[7-10,12] by exploiting room-temperature phase transitions with similar values of $|p|$; and the BC effects of $|\Delta T| \sim 9$ K that were achieved[50] away from a phase transition with a smaller value of $|p| = 0.18$ GPa in organic poly(methyl methacrylate) at $T_s \sim 368$ K.

To exploit our material in BC cooling devices, the non-monolithic working body and its intermixed pressure-transmitting medium may exchange heat with sinks and loads via fluid in a secondary circuit, heat pipes or fins[51]. The requisite high pressures could be generated in large volumes using small loads and small-area pistons, just as small voltages can generate large electric fields in the many thin films of an electrocaloric multilayer capacitor[52,53]. To improve the BC working body, it would be attractive to decrease the observed hysteresis using both chemical and physical approaches, enhance the limited thermal conductivity e.g. by two orders of magnitude via the introduction of graphite matrices[54], and combine different plastic crystals that operate at quite different temperatures[1,51,55]. More generally, our observation of colossal and reversible BC effects in NPG should inspire the study of BC effects in other mesophase systems that lie between liquids and solids, most immediately other organic plastic crystals whose PC-OC transitions display large latent heats and large volume changes[51,56].

**Experimental Section**

Samples:

NPG of purity of 99% was purchased as a powder from Sigma-Aldrich. The typical grain size was ~100 μm, as determined using optical microscopy.

Techniques:

Measurements of $dQ/|dT| = \frac{dQ/dt}{|dT/dt|}$ were performed at atmospheric pressure in a commercial TA Q100 differential scanning calorimeter (DSC), at 1-10 K min$^{-1}$, using ~ 10-20 mg samples of NPG ($t$ is time).

Measurements of specific heat $C_p$ were performed at atmospheric pressure in a commercial TA Q2000 DSC, at 5 K min$^{-1}$, using ~20 mg samples of NPG. Values of $C_p$ were obtained by recording heat flow out of/into the sample as a function of temperature, and comparing it with the heat flow out of/into a reference sapphire sample under the same conditions[57]. Latent heat $|Q_0| = \left|\int_{T_1}^{T_2} \frac{dQ}{dT} dT\right|$ across the PC-OC transition was obtained after subtracting baseline backgrounds, with start temperature $T_1$ freely chosen below (above) the transition on heating (cooling), and finish temperature $T_2$ freely chosen above (below) the transition on heating (cooling).

Measurements of $dQ/dT$ were performed at constant applied pressure using two bespoke differential thermal analysers (DTAs). For applied pressures of <0.3 GPa, we used a Cu-Be Bridgman pressure cell with chromel-alumel thermocouples. For applied pressures of <0.6 GPa, we used a model MV1-30 high-pressure cell (Institute of High Pressure Physics, Polish Academy of Science) with Peltier elements as thermal sensors. The temperature of both pressure cells was controlled using a circulating thermal bath (Lauda Proline RP 1290) that permitted the measurement temperature to be varied at ~2 K min$^{-1}$ in 183-473 K. NPG samples of mass ~100 mg were mixed with an inert perfluorinated liquid (Galden, Bioblock Scientist) to remove any residual air, and hermetically encapsulated inside Sn containers. The pressure-transmitting medium was DW-Therm (Huber Kältemaschinenbau GmbH). Entropy change $|\Delta S_0(p)| = \left| \int_{T_1}^{T_2} (dQ/dT)/T \, dT \right|$ across the PC-OC transition was obtained after subtracting baseline backgrounds, and the choice of $T_1$ and $T_2$ explained above.

Variable-temperature high-resolution x-ray diffraction was performed at atmospheric pressure in transmission, using Cu K$\alpha_1$ = 1.5406 Å radiation in a horizontally mounted INEL diffractometer with a quartz monochromator, a cylindrical position-sensitive detector (CPS-120) and the Debye-Scherrer geometry. NPG samples were introduced into a 0.5-mm-diameter Lindemann capillary to minimize absorption, and the temperature was varied using a 600 series Oxford Cryostream Cooler. Using the Materials Studio software[58], lattice parameters were determined by pattern matching using Pawley method for the cubic phase, and by Rietveld refinement for the monoclinic phase.

Dilatometry was performed using a bespoke apparatus that operated up to 0.3 GPa over ~193-433 K temperature range. Molten NPG samples of mass ~1 g were encapsulated inside stainless-steel containers to remove any residual air. Each container was then perforated by a stainless-steel piston, whose relative displacement with respect to a surrounding coil could be detected via measurement of electromotive force[59].

Variable-pressure x-ray diffraction measurements were performed at beamline MSPD BL04 in the ALBA-CELLS synchrotron[60], using an x-ray wavelength of 0.534 Å obtained at the Rh K-edge. The beamline is equipped with Kirkpatrick-Baez mirrors to focus the x-ray beam to 20 μm × 20 μm, and uses a Rayonix CCD detector. The NPG sample was placed with two small ruby chips at the centre of a 300 μm-diameter hole in a stainless steel gasket, preindented to a thickness of 55 μm. For room-temperature measurements, we used symmetric diamond-anvil cells (DACs) with diamonds of 700 μm. For high-temperature measurements, we used a gas-membrane driven DAC equipped with diamonds possessing 400 μm culets, and varied the temperature using a resistive heater. Temperature was measured using a K-type thermocouple attached to one diamond anvil, close to the gasket. The thermocouple was accurate to 0.4% in our measurement-set temperature range. For all the measurements, NaCl powder was used as the pressure marker[61]. The accuracy of pressure readings was ~±0.05GPa. Indexing and refinement of the powder patterns were performed using the Materials Studio software, by pattern matching using Pawley method.

Construction of entropy curves:

Using specific heat data at atmospheric pressure [Figure 1(b)], specific volume data at atmospheric pressure [Figure 1(d)], and $dQ/|dT|$ data at constant pressure [Figure 1(a), Figure 2(a,b)], we calculated $S'(T,p) = S(T,p) - S(250 \text{ K}, 0)$ using:

$$S'(T,p) = \begin{cases} \int_{250 \text{ K}}^{T_1} \dfrac{C_{\text{OC}}(T')}{T'} dT' + \Delta S_+(p) & T \leq T_1 \\ S(T_1, p) + \int_{T_1}^{T} \dfrac{1}{T'} \left( C_{\text{OC-PC}}(T') + \left| \dfrac{dQ(T',p)}{dT'} \right| \right) dT' + \Delta S_+(p) & T_1 \leq T \leq T_2 \\ S(T_2, p) + \int_{T_2}^{T} \dfrac{C_{\text{PC}}(T')}{T'} dT' + \Delta S_+(p) & T \geq T_2 \end{cases}$$

where $T_1$ is the transition start temperature, $T_2$ is the transition finish temperature, $C_{\text{OC}}$ is the specific heat of the OC phase outside the transition region, $C_{\text{PC}}$ is the specific heat of the PC phase outside the transition region, and $C_{\text{OC-PC}} = (1-x)C_{\text{OC}} + xC_{\text{PC}}$ represents the specific heat inside the transition region, where the transformed fraction $x$ on crossing the PC-OC transition was calculated using:

$$x = \left[ \int_{T_1}^{T} (dQ/dT')\,dT' \right] / \left[ \int_{T_1}^{T_2} (dQ/dT')\,dT' \right].$$

All values of specific heat are assumed to be independent of pressure.

**Acknowledgements**

This work was supported by the MINECO projects MAT2016-75823-R and FIS2017-82625-P, the DGU project 2017SGR-42, the UK EPSRC grant EP/M003752/1, and the ERC Starting grant no. 680032. We acknowledge ALBA for time on MSPD BL04 under proposal 2016021701. E. S.-T. and X. M. are grateful for support from the Royal Society.

| Compound | $T$ [K] | $|\Delta S|$ [J kg$^{-1}$ K$^{-1}$] | $|p|$ [GPa] | Reversible | Ref. |
|---|---|---|---|---|---|
| NPG | 320 | 445<br>510 | 0.25<br>0.52 | yes | This work |
| Ni$_{49.26}$Mn$_{36.08}$In$_{14.66}$ | 293 | 24 | 0.26 | partial | 7 |
| LaFe$_{11.35}$o$_{0.47}$Si$_{1.2}$ | 237 | 8.6 | 0.20 | partial | 8 |
| Gd$_5$Si$_2$Ge$_2$ | 270 | 11 | 0.20 | partial | 9 |
| Fe$_{49}$Rh$_{51}$ | 308 | 12 | 0.25 | partial | 10 |
| Mn$_3$GaN | 285 | 22 | 0.14 | partial | 11 |
| (MnNiSi)$_{0.62}$(FeCoGe)$_{0.38}$ | 330 | 70 | 0.27 | yes | 12 |
| BaTiO$_3$ | 400 | 1.6 | 0.10 | yes | 13 |
| (NH$_4$)$_2$SO$_4$ | 219 | 60 | 0.10 | yes | 14 |
| (NH$_4$)$_2$SnF$_6$ | 105 | 61 | 0.10 | yes | 15 |
| [TPrA]Mn[dca]$_3$ | 330 | 30 | 0.007 | yes | 16 |
| AgI | 390 | 60 | 0.25 | yes | 17 |
| Fluid R134a | 310 | 520 | 0.001 | yes | 18 |

**Table 1. BC effects near first-order phase transitions.** Isothermal entropy change $|\Delta S|$ at temperature $T$ due to changes of hydrostatic pressure $|p|$ (the nearby values of transition temperature $T_0$ appear in Table S1). All entries for BC solids denote data derived from quasi-direct measurements[30]. For the fluid hydrofluorocarbon R134a (1,1,1,2-tetrafluoroethane, i.e. CH$_2$FCF$_3$), the value of $|\Delta S|$ represents the full condensation of the fluid at 310 K and 0.001 MPa, while exploited in a typical vapour-compression refrigeration cycle[18].

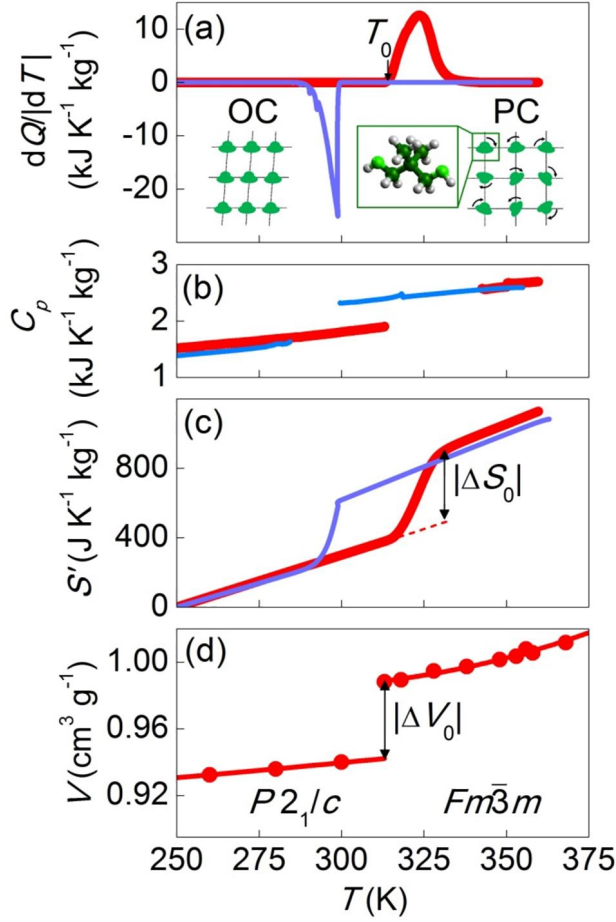

**Figure 1. Thermally driven phase transition in NPG at atmospheric pressure.** (a) Measurements of d$Q$/|d$T$| after baseline subtraction, on heating (red) and cooling (blue) across the first-order cubic-monoclinic phase transition, revealing a large latent heat. The insets represent simplified plan views of the globular $(CH_3)_2C(CH_2OH)_2$ molecules (C = ●, H = ● and O = ●), which are configurationally ordered in the monoclinic ordered-crystal (OC) phase (left inset), and configurationally disordered in the cubic plastic-crystal (PC) phase (right inset). We assume only one molecule per unit cell for ease of representation. (b) Specific heat $C_p$ either side of the transition on heating (red) and cooling (blue). (c) Entropy $S'(T) = S(T) - S(250\text{ K})$, evaluated via

$$S'(T) = S(T) - S(250\text{ K}) = \int_{250\text{K}}^{T} \left(C_p + |dQ/dT'|\right)/T' \, dT',$$

revealing a large entropy change $|\Delta S_0|$ for the transition. (d) Specific volume $V(T)$ on heating, revealing a large volume change $|\Delta V_0|$ for the transition. Symbols represent experimental data, lines are guides to the eye.

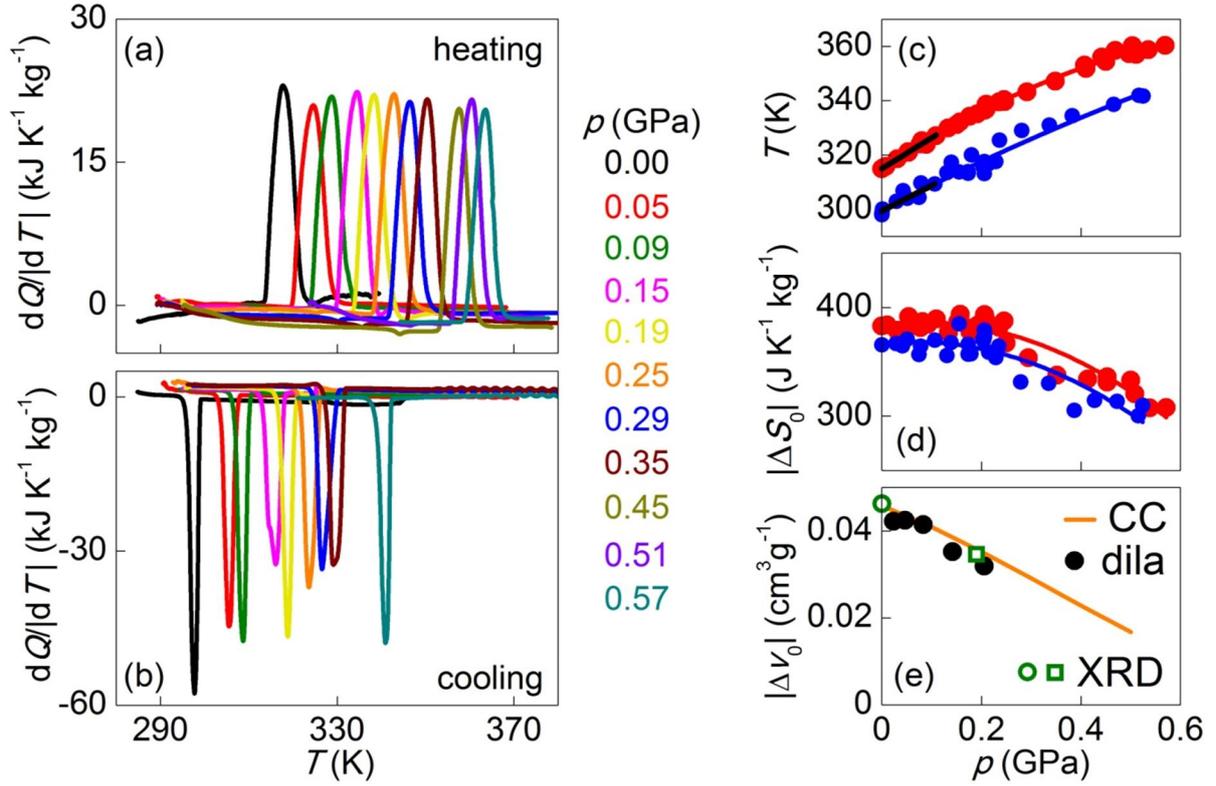

**Figure 2. Pressure-driven phase transition in NPG.** (a,b) Measurements of d$Q$/|d$T$| on heating and cooling across the first-order PC-OC transition for different values of increasing pressure $p$, after baseline subtraction. (c,d) Transition temperature and entropy change |$\Delta S_0(p)$| on heating (red symbols) and cooling (blue symbols), derived from the calorimetric data of (a,b) and equivalent data at other pressures (shown in Figure S2). Black lines in (c) are linear fits. Red and blue lines in (c,d) are guides to the eye. (e) Volume change for the transition |$\Delta V_0(p)$|: solid symbols obtained from the dilatometric data in Figure S3(a); open circle obtained from the x-ray diffraction data in Figure 1(c), open square obtained from the x-ray diffraction data in Figure S3(b); orange line obtained from (c,d) via the Clausius-Clapeyron (CC) equation.

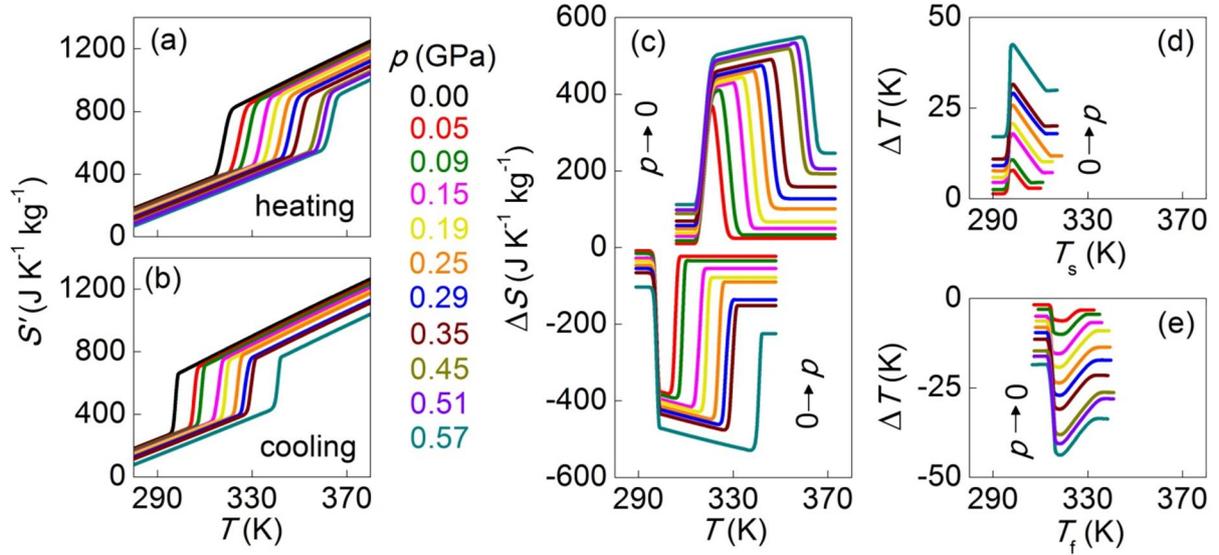

**Figure 3. Colossal BC effects in NPG near room temperature.** (a,b) Entropy $S'(T,p)$ with respect to the absolute entropy at 250 K and $p \sim 0$, on (a) heating and (b) cooling through the first-order PC-OC phase transition. (c) Isothermal entropy change $\Delta S$ for $0 \rightarrow p$ deduced from (b), and $p \rightarrow 0$ deduced from (a). (d) Adiabatic temperature change $\Delta T$ versus starting temperature $T_s$, for $0 \rightarrow p$ deduced from (b). (e) Adiabatic temperature change $\Delta T$ versus finishing temperature $T_f$ for $p \rightarrow 0$ deduced from (a).

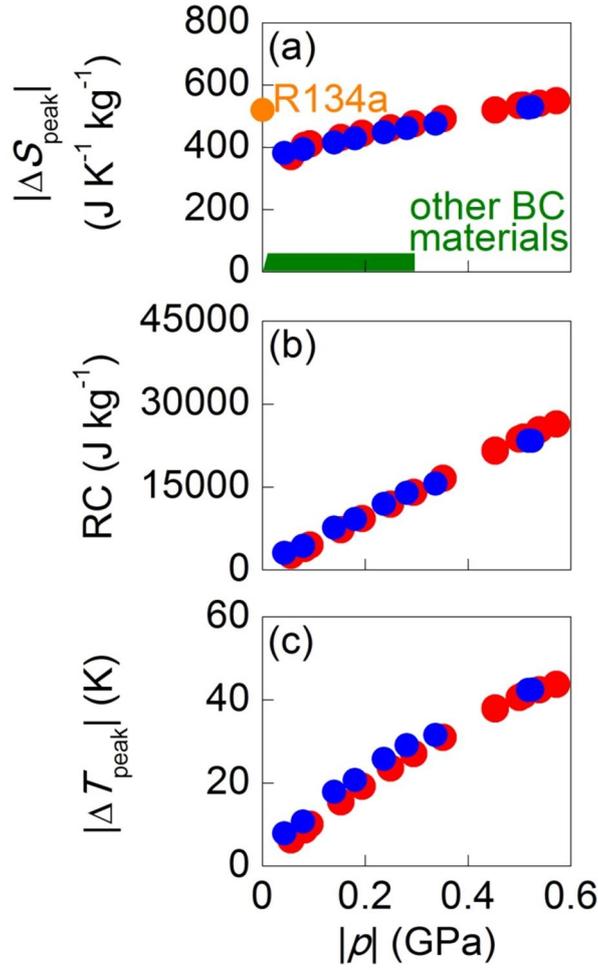

**Figure 4. BC performance near room temperature.** (a) For NPG, we show the peak isothermal entropy change $|\Delta S_{\text{peak}}|$ for pressure changes of magnitude $|p|$, on applying pressure (blue symbols) and removing pressure (red symbols). For comparison, the green envelope represents state-of-the-art BC materials (Table 1) that operate near room temperature, and the orange symbol represents the standard commercial fluid refrigerant[18] R134a for which operating pressures are ~0.001 GPa. For NPG alone, we show (b) refrigerant capacity RC = $|\Delta S_{\text{peak}}|$ × [FWHM of $\Delta S(T)$] and (c) peak values of the adiabatic temperature change $|\Delta T_{\text{peak}}|$, on applying pressure (blue symbols) and removing pressure (red symbols) near room temperature.

Supplementary Information

**Colossal barocaloric effects near room temperature in plastic crystals of neopentylglycol**

P. Lloveras[1], A. Aznar[1], M. Barrio[1], Ph. Negrier[2], C. Popescu[3], A. Planes[4], L. Mañosa[4], E. Stern-Taulats[5], A. Avramenko[5], N. D. Mathur[5], X. Moya[5,*] and J.-Ll. Tamarit[1,*]

[1]Grup de Caracterització de Materials, Departament de Física, EEBE and Barcelona Research Center in Multiscale Science and Engineering, Universitat Politècnica de Cataluna, Av. Eduard Maristany 10-14, 08019 Barcelona, Catalonia.

[2]Université de Bordeaux, LOMA, UMR 5798, F-33400 Talence, France.

[3]CELLS-ALBA, BP1413, 08290 Cerdanyola del Vallès, Catalonia.

[4]Facultat de Física, Departament de Física de la Matèria Condensada, Universitat de Barcelona, Martí i Franquès 1, 08028 Barcelona, Catalonia.

[5]Department of Materials Science, University of Cambridge, Cambridge, CB3 0FS, UK.

# Phase transition properties for first-order giant BC materials

| Compound | $T_0$ [K] | $|Q_0|$ [kJ kg$^{-1}$] | $|\Delta V_0/V_0|$ [%] | $|dT/dp|$ [K GPa$^{-1}$] | Ref. |
|---|---|---|---|---|---|
| NPG | 314 | 121 | 4.9 | 103 | This work |
| Ni$_{49.26}$Mn$_{36.08}$In$_{14.66}$ | 290 | 7.8 | 0.5 | 18 | S1 |
| LaFe$_{11.35}$o$_{0.47}$Si$_{1.2}$ | 250 | 2.8 | 1 | 94 | S2 |
| Gd$_5$Si$_2$Ge$_2$ | 260 | 5.5 | 1 | 35 | S3 |
| Fe$_{49}$Rh$_{51}$ | 319 | 3.8 | 1 | 59 | S4 |
| Mn$_3$GaN | 290 | 6.4 | 1 | 65 | S5 |
| (MnNiSi)$_{0.62}$(FeCoGe)$_{0.38}$ | 338 | 21 | 4 | 75 | S6 |
| BaTiO$_3$ | 400 | 0.9 | 0.1 | 55 | S7 |
| (NH$_4$)$_2$SnF$_6$ | 110 | 4.3 | 1 | 157 | S8 |
| (NH$_4$)$_2$SO$_4$ | 224 | 15 | 1 | 45 | S9 |
| [TPrA]Mn[dca]$_3$ | 330 | 14 | 1.25 | 231 | S10 |
| AgI | 420 | 27 | 5 | 134 | S11 |

**Table S1.** For the solids in Table 1, we show transition temperature $T_0$, latent heat $|Q_0|$, the relative specific volume change $|\Delta V_0|/V_0$, and the tunability of transition temperature with pressure $|dT/dp|$. For NPG, $|dT/dp|$ was obtained by averaging $dT/dp$ on heating and cooling for $p < 0.1$ GPa [Figure 2(c)].

**Thermally driven phase transition in NPG at different temperature ramp rates**

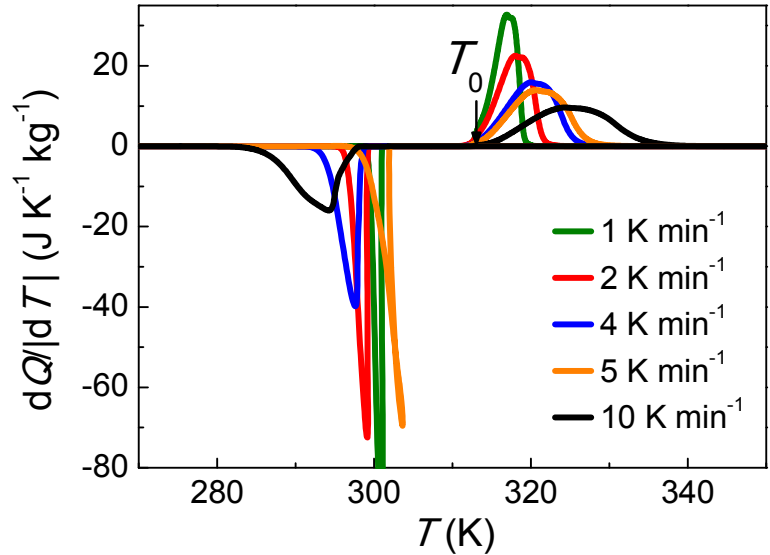

**Figure S1.** Measurements of d$Q$/|d$T$| after baseline subtraction, on heating (d$Q$/|d$T$| > 0) and cooling (d$Q$/|d$T$| < 0) across the first-order PC-OC phase transition, using different temperature ramp rates. The temperature ramp rate influences the transition finish temperature on heating, and the transition start and finish temperatures on cooling. By contrast, the temperature ramp rate has nominally no effect on $|Q_0| = \left| \int_{T_1}^{T_2} \frac{dQ}{dT} dT \right|$ across the PC-OC transition.

**Quasi-direct measurements performed using the Cu-Be Bridgman pressure cell**

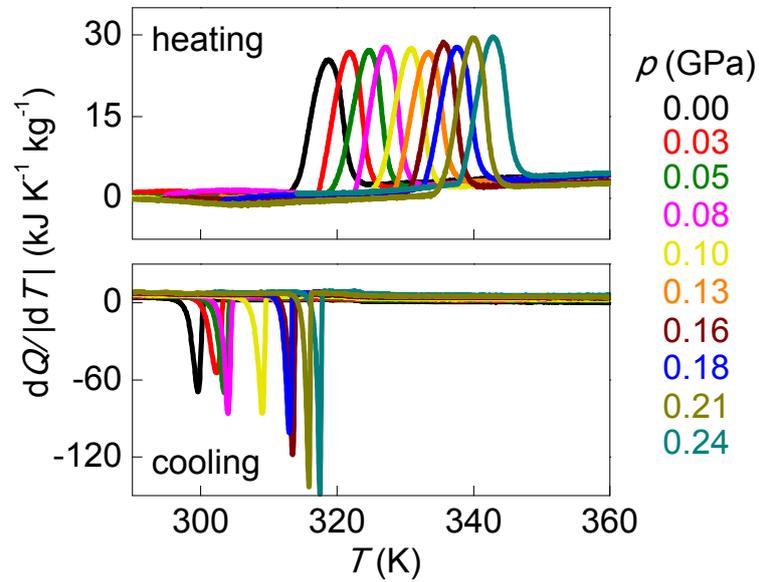

**Figure S2.** Measurements of $dQ/|dT|$ on heating ($dQ/|dT| > 0$) and cooling ($dQ/|dT| < 0$) across the first-order PC-OC transition for different values of increasing pressure $p$, after baseline subtraction, at ~2 K min$^{-1}$. Data taken using the Cu-Be Bridgman pressure cell with chromel alumel thermocouples.

**Pressure-dependent changes in volume across the transition**

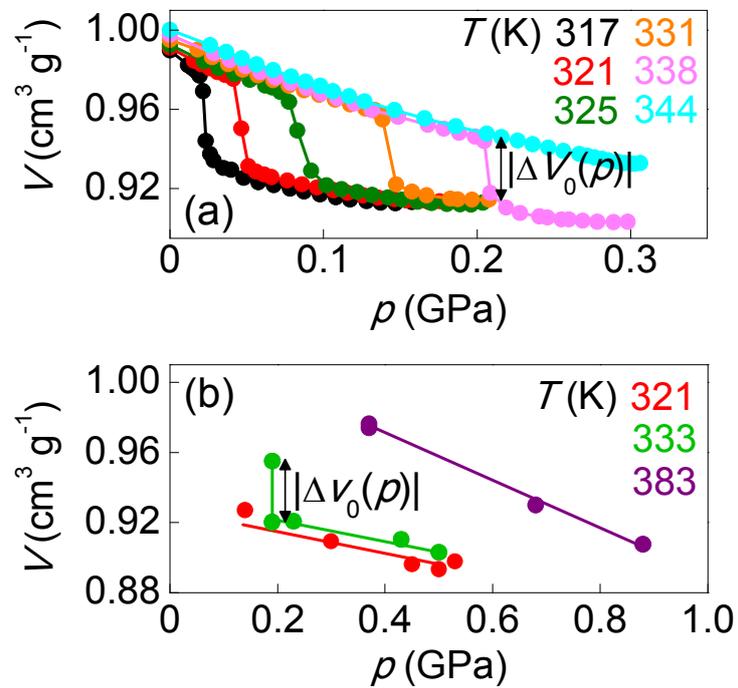

**Figure S3.** (a) Pressure-dependent specific volume $V$, obtained by dilatometry while isothermally removing an applied pressure of 0.3 GPa at different measurement temperatures. (b) Pressure-dependent specific volume $V$, obtained by x-ray diffraction (XRD) while isothermally removing an applied pressure of 0.9 GPa at different measurement temperatures. Measurement errors in pressure are ±0.002 GPa for dilatometry, and ±0.05 GPa for high-pressure x-ray diffraction.

**Coefficient of isobaric thermal expansion**

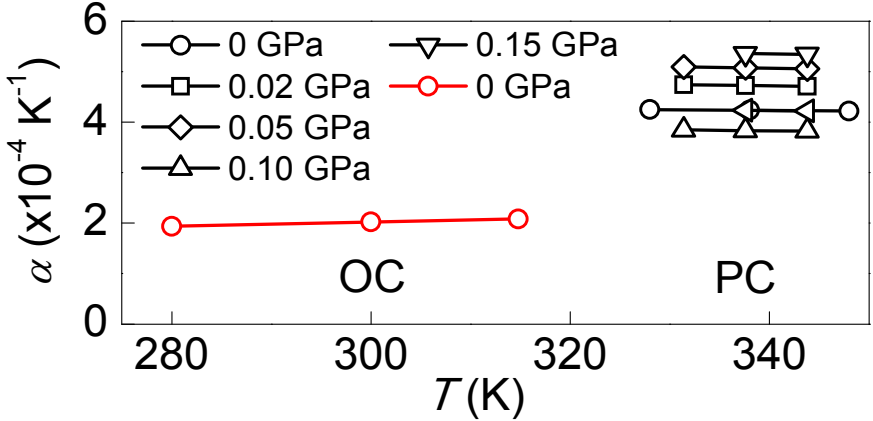

**Figure S4.** Coefficient of isobaric thermal expansion $\alpha$ for the PC phase (black) and the OC phase (red), near the PC-OC transition, obtained from data in Figure 1(d) and Figure S3(a). The non-monotonic changes of $\alpha$ with pressure for the PC phase imply a ~20% error in $(\partial V/\partial T)_p$.

**Coefficient of isothermal compressibility**

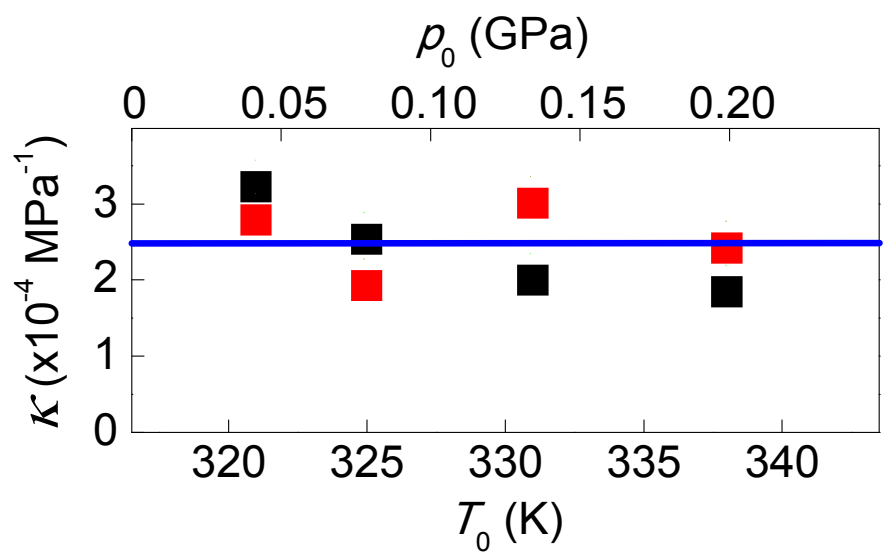

**Figure S5.** Coefficient of isothermal compressibility $\kappa$ for the PC phase (black symbols) and the OC phase (red symbols), near the PC-OC transition, obtained from data in Figure S3(a). The blue line represents the average value of $\kappa$ for all data.

**Temperature-dependent x-ray diffraction**

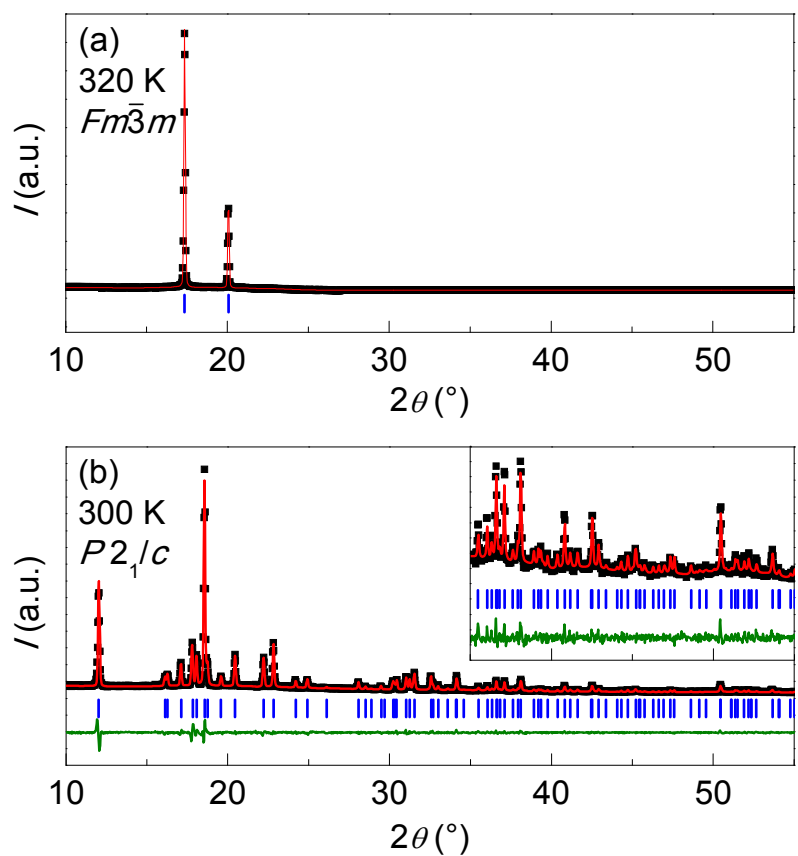

**Figure S6.** Detail of selected x-ray diffraction spectra obtained (a) above and (b) below the PC-OC transition on heating, at atmospheric pressure. Black symbols are experimental data, red lines are fitted patterns, vertical blue lines indicate indexed reflections (shown overleaf).

| Indexed reflections | | | | | |
|---|---|---|---|---|---|
| T = 300 K | | | | T = 320 K | |
| 2θ (°) | hkl | 2θ (°) | Hkl | 2θ (°) | hkl |
| 12.03 | 0 1 1 | 36.34 | -1 4 1 | 17.35 | 1 1 1 |
| 16.12 | 1 0 0 | 36.60 | 2 2 0 | 20.06 | 2 0 0 |
| 16.26 | 0 2 0 | 36.65 | 0 3 3 | | |
| 17.11 | -1 1 1 | 36.82 | 1 4 0 | | |
| 17.82 | 0 0 2 | 37.11 | -2 2 3 | | |
| 18.08 | 1 1 0 | 37.63 | 0 4 2 | | |
| 18.57 | 0 2 1 | 38.12 | 2 1 1 | | |
| 18.75 | -1 0 2 | 37.94 | 1 3 2 | | |
| 19.59 | 0 1 2 | 38.93 | -2 1 4 | | |
| 20.46 | -1 1 2 | 39.20 | -2 3 1 | | |
| 22.23 | -1 2 1 | 39.36 | -2 3 2 | | |
| 22.84 | 1 1 1 | 39.77 | 0 2 4 | | |
| 24.20 | 0 2 2 | 40.37 | 1 2 3 | | |
| 24.91 | -1 2 2 | 40.83 | 2 2 1 | | |
| 26.11 | 0 3 1 | 41.63 | -2 3 3 | | |
| 28.08 | 0 1 3 | 42.46 | 0 5 1 | | |
| 28.52 | 1 0 2 | 42.54 | -1 1 5 | | |
| 28.87 | -1 3 1 | 42.93 | 0 4 3 | | |
| 29.49 | 1 3 0 | 43.39 | 2 0 2 | | |
| 29.69 | 1 1 2 | 44.07 | 0 3 4 | | |
| 30.23 | -1 2 3 | 44.31 | -1 5 1 | | |
| 30.32 | -2 0 2 | 44.73 | 1 5 0 | | |
| 30.43 | 0 3 2 | 45.22 | -2 1 5 | | |
| 31.03 | -1 3 2 | 45.42 | 0 5 2 | | |
| 31.25 | -2 1 1 | 45.49 | 1 0 4 | | |
| 31.54 | 0 2 3 | 45.75 | -2 3 4 | | |
| 32.57 | 2 0 0 | 46.29 | 1 1 4 | | |
| 32.69 | 1 3 1 | 46.63 | 2 2 2 | | |
| 33.00 | 1 2 2 | 46.95 | 2 4 0 | | |
| 33.60 | 2 1 0 | 47.39 | -1 4 4 | | |
| 34.12 | 0 4 1 | 47.60 | -2 2 5 | | |
| 34.15 | -1 1 4 | 48.63 | 0 2 5 | | |
| 34.59 | -2 2 2 | 49.16 | -1 5 3 | | |
| 35.48 | -1 3 3 | 49.57 | 0 4 4 | | |
| 36.04 | 0 0 4 | | | | |

**Table S3.** Indexed reflections for the the x-ray diffraction patterns shown in Figure S4.

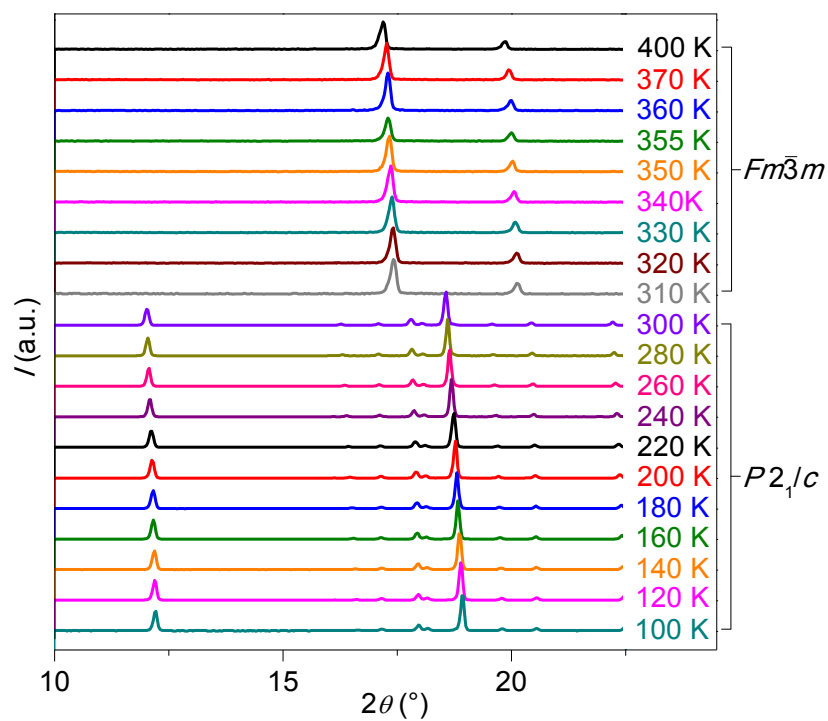

**Figure S7.** Detail of all x-ray diffraction spectra obtained across the PC-OC transition on heating, at atmospheric pressure.

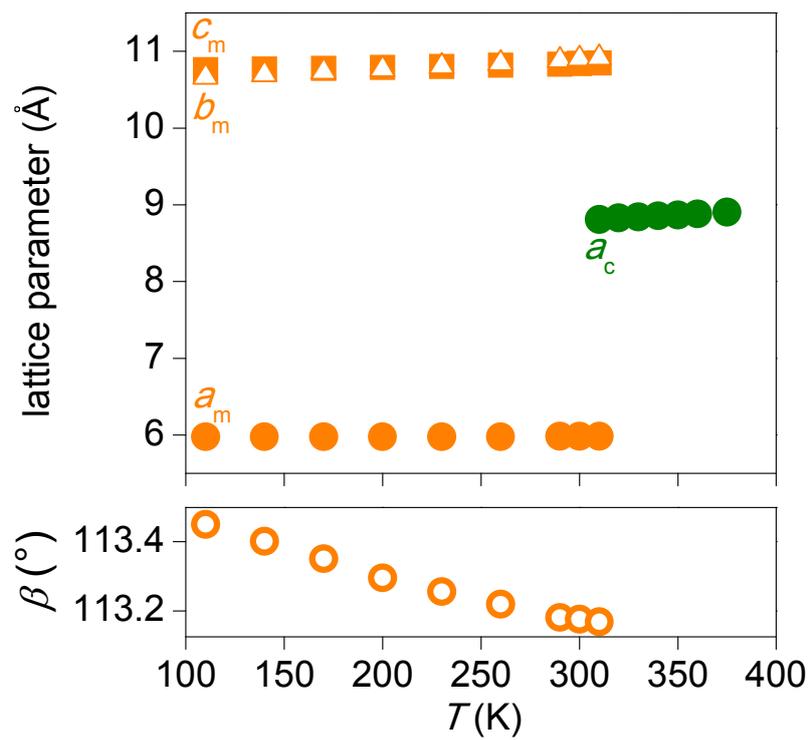

**Figure S8.** Temperature dependence of lattice parameters across the PC-OC phase transition, obtained from x-ray diffraction on heating.

**Pressure-dependent x-ray diffraction**

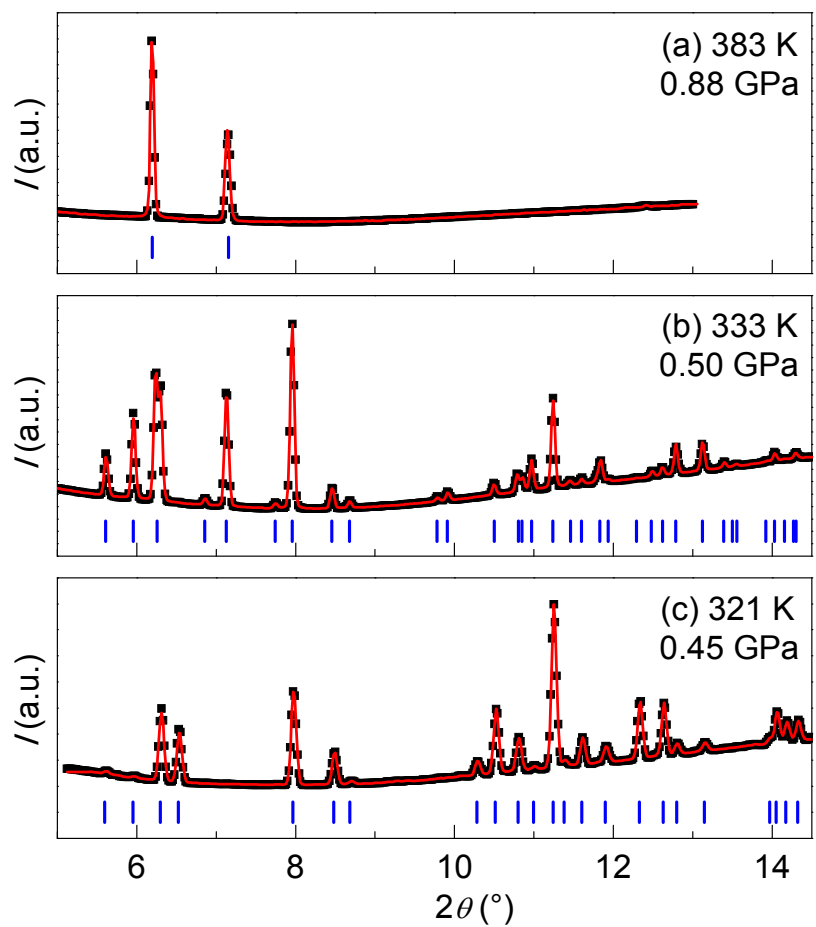

**Figure S9.** Detail of selected x-ray diffraction spectra obtained under hydrostatic pressure. Black symbols are experimental data, red lines are fitted patterns, vertical blue lines indicate indexed reflections (shown overleaf).

| Indexed reflections | | | | | |
|---|---|---|---|---|---|
| T = 321 K, p = 0.45 GPa | | T = 333 K, p = 0.50 GPa | | T = 383 K, p = 0.88 GPa | |
| 2θ (°) | hkl | 2θ (°) | hkl | 2θ (°) | hkl |
| 5.59 | 1 0 0 | 5.61 | 1 0 0 | 6.19 | 1 1 1 |
| 5.95 | -1 1 1 | 5.96 | -1 1 1 | 7.15 | 2 0 0 |
| 6.30 | 1 1 0 | 6.26 | 0 0 2 | | |
| 6.52 | 0 2 1 | 6.86 | 0 1 2 | | |
| 6.52 | -1 0 2 | 7.13 | -1 1 2 | | |
| 7.96 | 1 1 1 | 7.74 | -1 2 1 | | |
| 8.48 | 0 2 2 | 7.95 | 1 1 1 | | |
| 8.69 | -1 2 2 | 8.46 | 0 2 2 | | |
| 10.28 | 1 3 0 | 8.68 | -1 2 2 | | |
| 10.52 | -1 2 3 | 9.78 | 0 1 3 | | |
| 10.80 | -2 1 1 | 9.91 | 1 0 2 | | |
| 11.00 | 0 2 3 | 10.50 | -1 2 3 | | |
| 11.24 | 2 0 0 | 10.81 | -2 1 1 | | |
| 11.38 | 1 3 1 | 10.85 | -2 1 2 | | |
| 11.60 | 2 1 0 | 10.97 | 0 2 3 | | |
| 11.90 | 0 4 1 | 11.24 | 2 0 0 | | |
| 12.33 | -1 3 3 | 11.46 | 1 2 2 | | |
| 12.63 | -1 4 1 | 11.60 | 2 1 0 | | |
| 12.63 | 2 2 0 | 11.83 | -1 1 4 | | |
| 12.80 | 1 4 0 | 11.94 | -2 2 2 | | |
| 12.80 | -2 2 3 | 12.29 | -1 3 3 | | |
| 13.15 | 2 1 1 | 12.48 | 0 0 4 | | |
| 13.97 | 1 2 3 | 12.62 | 2 2 0 | | |
| 14.05 | 2 2 1 | 13.92 | 1 2 3 | | |
| 14.17 | 2 3 0 | 14.03 | 2 2 1 | | |
| 14.32 | -2 3 3 | 14.15 | 2 3 0 | | |
| | | 14.27 | -2 2 4 | | |
| | | 14.30 | -2 3 3 | | |

**Table S3.** Indexed reflections for the x-ray diffraction patterns shown in Figure S7.

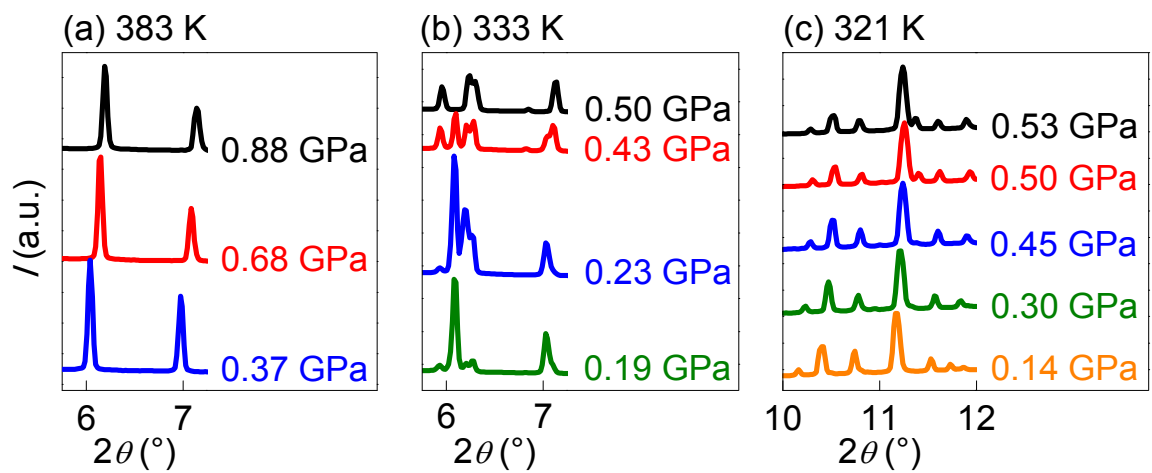

**Figure S10.** Detail of x-ray diffraction spectra obtained under hydrostatic pressure, at various temperatures.

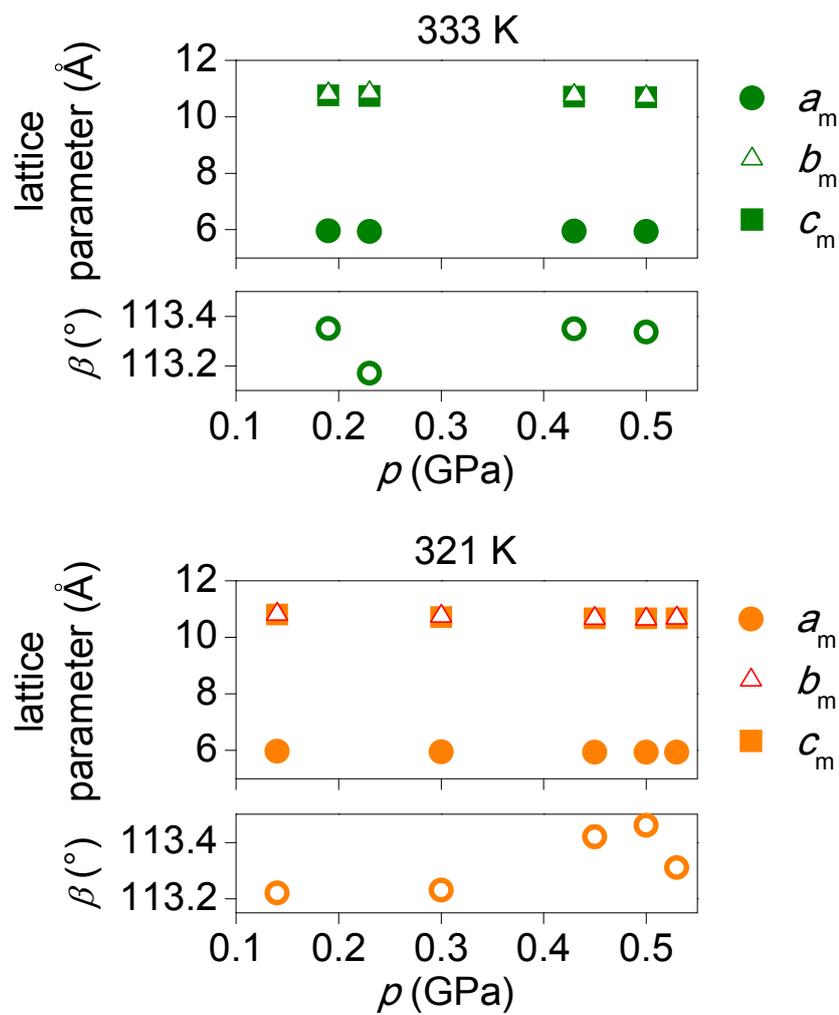

**Figure S11.** Pressure dependence of lattice parameters in the OC phase, obtained from x-ray diffraction under hydrostatic pressure.